\documentstyle[11pt,aaspp4]{article}

\newcommand\etal{et al.}

\newcommand\tg{\left( \begin{array}{cc} 
g_{a} & 0 \\
0 & g_{b} 
\end{array}\right)}
\newcommand\sigmax{\left( \begin{array}{cc}
0 & 1 \\
1 & 0 
\end{array}\right)}
\newcommand\sigmay{\left( \begin{array}{cc}
0 & -i \\
i & 0 
\end{array}\right)}
\newcommand\sigmaz{\left( \begin{array}{cc}
1 & 0 \\
0 & -1 
\end{array}\right)}
\newcommand\cmatrix{\left( \begin{array}{cc} 
\cos\theta_{a}\cos\chi_{a} + i\sin\theta_{a}\sin\chi_{a} & 
-\sin\theta_{a}\cos\chi_{a} + i\cos\theta_{a}\sin\chi_{a} \\ 
\sin\theta_{b}\cos\chi_{b} + i\cos\theta_{b}\sin\chi_{b} & 
\cos\theta_{b}\cos\chi_{b} -  i\sin\theta_{b}\sin\chi_{b} 
\end{array}
\right)}

\lefthead{BRITTON}
\righthead{RADIO ASTRONOMICAL POLARIMETRY AND THE LORENTZ GROUP}

\begin{document}

\title{Radio Astronomical Polarimetry and the Lorentz Group}

\author{M. C. Britton}

\affil{Swinburne Centre for Astrophysics and Supercomputing, 
Swinburne University of Technology, Hawthorn, Victoria 3122, Australia}
\authoraddr{mail communications to 
        Matthew Britton 
	Astrophysics and Supercomputing Group
	Mail \# 31
        Swinburne University                   phone: +61-3-9214 5622
        PO Box 218                             fax: +61-3-9819 0856
        Hawthorn Victoria 3122
        Australia
or email mbritton@mania.physics.swin.edu.au}

\abstract{In radio astronomy the polarimetric properties of radiation
are often modified during propagation and reception.  Effects such as
Faraday rotation, receiver cross-talk, and differential amplification
act to change the state of polarized radiation.  A general description
of such transformations is useful for the investigation of these
effects and for the interpretation and calibration of polarimetric
observations.  Such a description is provided by the Lorentz group,
which is intimately related to the transformation properties of
polarized radiation.  In this paper the transformations that commonly
arise in radio astronomy are analyzed in the context of this group.
This analysis is then used to construct a model for the propagation
and reception of radio waves.  The implications of this model for
radio astronomical polarimetry are discussed.}


\section{Introduction}

In radio astronomy transformations occur during the propagation and
reception of radio waves that act to change the state of polarized
radiation.  Some of these transformations, such as Faraday rotation in
the ionosphere or the interstellar medium, arise from propagation
effects that may themselves be of astrophysical interest.  Others
originate from instrumental effects such as differential amplification or
receiver cross-talk, and have an adverse effect on polarimetric
observations.  Realistically many such effects may be present, each
having its own time and frequency dependence, and collectively acting
to distort measurements of the polarized radiation.  The
interpretation and calibration of these observations may be quite
complex, and it is useful to have a general context in which to
describe these transformations.  

Linear transformations of fully polarized radiation were first
investigated by Jones (1941\markcite{jones}), who represented
transformations of the two-component transverse electric field in
terms of 2x2 complex matrices now called Jones matrices.  This
analysis was extended to partially polarized radiation by Parrent \&
Roman (1960\markcite{parrent}), who used Jones matrices to describe
the transformation properties of the coherency matrix.  Alternatively,
both fully and partially polarized radiation may be described by the
Stokes parameters, and their linear transformations may be represented
in terms of 4x4 real matrices called Mueller matrices (Mueller
1948\markcite{mueller}).

In fact these transformations are intimately related to the Lorentz
group.  This relationship arises from the fact that for a plane
propagating wave Maxwell's equations admit two independent solutions,
representing the two orthogonal senses of polarization.  Thus
polarized radiation constitutes a two state system, and the linear
transformations of all such systems are described by the Lorentz
group.  This relationship has long been known in optics, and different
aspects have been discussed by many authors (e.g. Barakat
1963\markcite{barakat}, Whitney 1971\markcite{whitney}, Cloude
1986\markcite{cloude}, Pellat-Finet \& Bausset 1992\markcite{pfb}).
Recent work has focused on representations of the Jones and Mueller
matrices (Opatrny \& Perina 1993\markcite{opat}, Brown \& Bak
1995\markcite{BB}, Han, Kim \& Noz 1997\markcite{hkna}).  Up to a
multiplicative constant the set of Jones matrices constitute the group
SL(2,C), which forms the spin 1/2 representation of the Lorentz group.
The corresponding set of Mueller matrices constitutes the group
SO(3,1) and forms the spin 1 representation of the Lorentz group.
Finally, the relationship between Jones and Mueller matrices is
represented through the 2-1 mapping between these two groups.
SO(3,1).

The formulation of the transformation properties of polarized
radiation in terms of Lorentz transformations affords considerable
insight, including the interpretation of the Stokes parameters as a
Lorentz 4-vector, the classification of transformations of this
4-vector as rotations or boosts, and the existence of a polarimetric
analogue to the invariant interval.  In this paper these concepts are
reviewed in the context of radio astronomical polarimetry.  Since both
linear and circular bases are widely used in radio astronomy, a basis
independent formulation is emphasized.  This formalism is then used to
construct a model for the propagation and reception of radio waves.

\section{Representations of Polarized Radiation}
\label{ereps}

Consider the representation of a transverse electromagnetic wave.
Such a wave may be described through its two-component transverse
electric field vector ${\bf{\cal E}}(t)$.  This vector may also
represent the electric field in a waveguide, or the voltage in a pair
of cables.  The vector ${\bf{\cal E}}(t)$ is commonly written in terms
of the two-component complex analytic signal ${\bf E}(t)$ as
${\bf{\cal E}}(t)={\rm Re}\left[ {\bf E}(t) \exp\{i\omega t\}\right]$.
This construction is familiar from both optics (Born \& Wolf
1980\markcite{born}) and signal processing (Bracewell
1986\markcite{brace}).

For fully polarized radiation the analytic signal is independent of
time.  In this case the relative amplitudes and phases of the two
components of ${\bf E}$ specify the state of elliptical polarization
of the plane wave.  For partially polarized radiation ${\bf E}(t)$ is
time dependent, and measurable properties of the wave may instead be
defined through time-averaging.  Such an averaging procedure is
conveniently treated through the coherency matrix (Wiener
1930\markcite{wienera}, Wolf 1959\markcite{wolf}).  This is a 2x2
Hermitian matrix formed from the direct product of the analytic
signal, and may be written as ${\bf \rho}= \langle{\bf E}(t){\bf
\otimes} {\bf E}^{\dagger}(t)\rangle$.  Here the angular brackets
denote time-averaging.  As with any Hermitian matrix, the coherency
matrix may be written in terms of 4 real quantities $(S_{o},{\bf
S})$ as
\begin{equation}
{\bf \rho} = (S_{o} \sigma_{o} + {\bf S} \cdot {\bf \sigma})/2
\label{dens}
\end{equation}
where $\bf \sigma_{o}$ is the 2x2 identity matrix and ${\bf \sigma}$
is a 3-vector whose components are the Pauli spin matrices.  These
three 2x2 matrices are traceless and Hermitian with determinant $-$1.
The 4 parameters $(S_{o},{\bf S})$ are simply the mean Stokes
parameters of the plane wave (Fano 1954\markcite{fano}), with $S_{o}$
representing the total intensity.

Let us now introduce a particular basis.  The electric field vector
may be represented in the Cartesian basis $(\hat{x}, \hat{y},
\hat{z})$, in which ${\cal E}(t) = ({\cal E}_{x}(t),{\cal E}_{y}(t))$
is resolved into mutually orthogonal components, each orthogonal to
the direction of propagation $z$ of the plane wave.  The corresponding
analytic signal is ${\bf E}(t) = (E_{x}(t),E_{y}(t))$.  For the
3-dimensional space of ${\bf S}$ the Cartesian basis $(\hat{q},
\hat{u}, \hat{v})$ is used, along with the customary representation of
the Pauli matrices
\begin{eqnarray}
{\bf \sigma}_{\hat{q}} = \sigmaz &
{\bf \sigma}_{\hat{u}} = \sigmax & 
{\bf \sigma}_{\hat{v}} = \sigmay 
\label{pauli}
\end{eqnarray}
To associate the coherency matrix with the Stokes parameters, we write
${\bf \rho}$ in this basis as
\begin{equation}
{\bf \rho} = \langle\bf{E}(t){\bf \otimes} \bf{E}^{\dagger}(t)\rangle 
  = \left( \begin{array}{cc}
	\langle E_{x}^{*}(t)E_{x}(t)\rangle & \langle E_{x}^{*}(t)E_{y}(t)\rangle \\
	\langle E_{y}^{*}(t)E_{x}(t)\rangle & \langle E_{y}^{*}(t)E_{y}(t)\rangle
	\end{array} \right) 
\end{equation}
From equation \ref{dens}, the Stokes parameters in this basis become
\begin{equation}
\begin{array}{cc}
\begin{array}{l}
S_{o} = \langle E_{x}^{*}(t)E_{x}(t)\rangle + \langle E_{y}^{*}(t)E_{y}(t)\rangle \\
S_{q} = \langle E_{x}^{*}(t)E_{x}(t)\rangle - \langle E_{y}^{*}(t)E_{y}(t)\rangle 
\end{array}
\begin{array}{l}
S_{u} = 2\;{\rm Re}\left[\langle E_{x}^{*}(t)E_{y}(t)\rangle \right] \\
S_{v} = 2\;{\rm Im}\left[\langle E_{x}^{*}(t)E_{y}(t)\rangle \right] 
\end{array}
\end{array}
\label{stokesdef}
\end{equation}
This is simply the usual definition of the Stokes parameters in a
linear basis (Born \& Wolf 1980\markcite{born}).  

\section{Transformation Properties of Polarized Radiation}
\label{xforms}

Let us now consider the transformation properties of polarized
radiation.  Attention is restricted to linear, invertible
transformations.  This excludes the class of projective
transformations, which are important in representing perfect
polarizing filters.  Similarly, such transformations cannot describe
multipath propagation of coherent radiation, such as the focusing or
defocusing of radiation by lenses or mirrors.  Despite this
restriction, the set of linear, invertible transformations encompasses
a broad class of physical processes, including single-particle
scattering, propagation through anisotropic media, and many
transformations arising from instrumental devices.  This set may also
describe the linear transformations of the voltage signal in two
cables, which are known in linear network theory as two-port networks
(Ruston \& Bordogna 1966\markcite{linnets}).  This equivalence is
particularly useful in radio astronomy, where the two components of
the electric field are converted to voltages by a receiver and then
passed through an electronics downconversion chain.

The most general linear transformation of the analytic signal may be
written as ${\bf E}'(t) = {\bf t}{\bf E}(t)$, where the Jones matrix
{\bf t} is a 2x2 complex matrix.  As the direct product of the
analytic signal, the coherency matrix must transform as ${\bf \rho}' =
{\bf t} {\bf \rho}\; {\bf t}^{\dagger}$.  For invertible
transformations a Jones matrix may be written as ${\bf t} =
\sqrt{{\rm det}\;{\bf t}}\;{\bf t}_{N}$, where ${\rm det}\;{\bf t}$ is
the determinant of ${\bf t}$ and ${\bf t}_{N}$ is a matrix with unit
determinant.  The set of 2x2 complex invertible matrices with unit
determinant forms the group SL(2,C), which constitutes the spin 1/2
representation of the Lorentz group.

To investigate the transformation properties of the Stokes parameters,
note that the determinant of equation \ref{dens} is simply ${\rm det}
\;{\bf \rho} = S_{o}^{2} - |{S}|^{2} \equiv S_{\rm inv}$ (Barakat
1963\markcite{barakat}).  This is just the form of the Lorentz
invariant.  Under transformation by the Jones matrix ${\bf t}$, ${\rm
det}\; {\bf \rho}' = |\sqrt{{\rm det}\;{\bf t}}|^{2} S_{\rm inv}$,
so that this interval is preserved up to a multiplicative constant.
The set of transformations that preserve this interval forms the group
SO(3,1), which constitutes the spin 1 representation of the Lorentz
group.  That is, the Stokes parameters transform as a Lorentz
4-vector, with the total intensity acting as the timelike component
and the remaining Stokes parameters acting as the spacelike
components.  The condition that the total intensity $S_{o}>0$
restricts this 4-vector to lie within or on the surface of the forward
light cone.  These two cases correspond to partially polarized
($S_{\rm inv}>0$) or fully polarized ($S_{\rm inv} = 0$) radiation,
respectively.

The representations of the groups SL(2,C) and SO(3,1) are well known
in physics, but are not widely used in astronomy.  Basis-independent
representations of these groups are now reviewed, and are interpreted
in the context of polarimetry.  This will serve both as an
introduction and to establish the notation used in the next section.
For similar reviews, see Brown \& Bak (1995\markcite{BB}) or Tung
(1996\markcite{Tung}).

The group SL(2,C) contains as a subgroup the set of 2x2 unitary
transformations SU(2).  Any such unitary transformation may be
parameterized as
\begin{equation}
{\bf r}_{\hat{\bf n}}(\phi) = 
e^{\left(i\,{\bf \sigma}\cdot\hat{{\bf n}}\,\phi\right)} = 
\sigma_{0}\,\cos \phi + i\,{\bf \sigma}\cdot\hat{{\bf n}}\,\sin \phi
\label{erot}
\end{equation}
where $\hat {\bf n}$ is a unit 3-vector.  This is called the
axis-angle parameterization of SU(2).  The angle $\phi$ differs from
the definition customary in classical and quantum mechanics by a factor
of $1/2$, but is in agreement with the conventions of optics.  Under
the transformation of equation \ref{erot}, ${\bf \rho} \rightarrow
{\bf r}_{\hat{\bf n}}(\phi) {\bf \rho}\; {\bf r}_{\hat{\bf n}}(-\phi)
= (\sigma_{o} S_{o} + {\bf \sigma}\cdot{\bf S}')/2$, where
\begin{equation}
{\bf S}' = {\bf S}\cos 2\phi + {\bf S}\times\hat{\bf n}\sin 2\phi
+ (\hat{\bf n}\cdot{\bf S})\hat{\bf n}\left(1-\cos 2\phi\right)
\label{vecrot}
\end{equation}
and we have used the relationship $({\bf \sigma}\cdot{\bf
a})({\bf \sigma}\cdot{\bf b}) = {\bf a}\cdot{\bf b}
+ i{\bf \sigma}\cdot({\bf a}\times{\bf b})$.  But ${\bf
{S}}'$ is simply the vector resulting from a rotation of ${\bf
S}$ about an axis $\hat{\bf n}$ by an angle $2\phi$ ({\it cf}.
Goldstein 1980\markcite{gstein}).  This reflects the well-known
mapping between SU(2) and the group SO(3), whose elements form a
representation of the rotations of a 3-dimensional vector.  The
mapping is 2-1, since the two rotations ${\bf r}_{\hat{\bf n}}(\phi)$
and ${\bf r}_{\hat{\bf n}}(\phi+\pi)= -{\bf r}_{\pm\hat{\bf n}}(\phi)$
result in the same vector ${\bf S}'$.  Such rotations preserve the
degree of polarization $|{\bf S}|/S_{o}$ of the plane wave, and
are readily interpreted geometrically in the space of the Poincare
sphere in terms of the axis $\hat{\bf n}$ and angle $2\phi$ of
rotation.  Equation \ref{erot} may be represented in the basis of
equation \ref{pauli} as
\begin{equation}
\bf{r}_{\hat{n}}(\phi) = \left(\begin{array}{cc}
              \cos \phi + in_{q}\sin \phi & (in_{u}+n_{v}) \sin \phi \\
              (in_{u}-n_{v}) \sin \phi  & \cos \phi -in_{q}\sin\phi
              \end{array} \right)
\label{erotbasis}
\end{equation}
where ${\hat {\bf n}} =(n_{\hat q}, n_{\hat u}, n_{\hat v})$.  From
equation \ref{vecrot} the corresponding rotation of the Stokes
parameters $(S_{o},S_{q},S_{u},S_{v})$ in this basis is
\begin{equation}
{\bf R}_{\hat{\bf n}}(2\phi) = \left(\begin{array}{cccc}
1 & 0 & 0 & 0 \\
0 & n_{q}^{2} + (1-n_{q}^{2})\cos 2\phi&n_{q}n_{u}\sin^{2}\phi+n_{v}\sin 2\phi&
	n_{q}n_{v}\sin^{2}\phi-n_{u}\sin 2\phi \\	
0 & n_{q}n_{u}\sin^{2}\phi-n_{v}\sin 2\phi&n_{u}^{2} + (1-n_{u}^{2})\cos 2\phi&
	n_{u}n_{v}\sin^{2}\phi+n_{q}\sin 2\phi \\
0 & n_{q}n_{v}\sin^{2}\phi+n_{u}\sin 2\phi & n_{u}n_{v}\sin^{2}\phi-n_{q}\sin 2\phi&
	n_{v}^{2} + (1-n_{v}^{2})\cos 2\phi
\end{array}\right)  
\label{srotbasis}
\end{equation}
For example, for $(n_{\hat q}, n_{\hat u}, n_{\hat v}) = (1,0,0)$
equation \ref{srotbasis} constitutes the rotation about the ${\hat q}$
axis ${\bf R}_{\hat{\bf q}}(2\phi)$, which may be interpreted from
equation \ref{erotbasis} as generating a phase delay between the two
components of the electric field.

Next let us consider the group SL(2,C), which has a parameterization
similar to its subgroup SU(2).  Any element of the group may be
written as $\exp\left(i {\bf \sigma} \cdot \hat{{\bf n}}\,\phi +
{\bf \sigma} \cdot \hat{{\bf m}}\,\beta\right)$, where ${\hat {\bf
n}}$ and ${\hat {\bf m}}$ are unit vectors.  In analogy with equation
\ref{erot}, let us consider transformations for which $\phi=0$.  Such
transformations may be written in terms of the Hermitian matrix
\begin{equation}
{\bf b}_{\hat{\bf m}}(\beta) = 
\exp\left({\bf \sigma}\cdot\hat{{\bf m}}\,\beta\right) = 
\sigma_{0}\,\cosh \beta + {\bf \sigma}\cdot\hat{{\bf m}}\,\sinh \beta
\label{eboost}
\end{equation}
Under this transformation the coherency matrix becomes ${\bf \rho}
\rightarrow {\bf b}_{\hat{\bf m}}(\beta) {\bf \rho}\; {\bf
b}_{\hat{\bf m}}(\beta) = (\sigma_{o}S_{o}' + {\bf
\sigma}\cdot{\bf S}')/2$, where
\begin{equation}
\begin{array}{c}
S_{o}' = S_{o}\cosh 2\beta + {\bf S}\cdot{\hat{\bf m}}\sinh 2\beta \\
{\bf S}' = {\bf S} + \left(S_{o}\sinh 2\beta + 
	2{\bf S}\cdot{\hat{\bf m}}\sinh^{2}\beta\right)\hat{\bf m}
\end{array}
\label{sboost}
\end{equation}
This is simply the result of performing a Lorentz boost on the
4-vector $S=(S_{o},{\bf S})$ along the axis $\hat{\bf m}$ by a
velocity parameter $2 \beta$ ({\it cf}. Jackson
1975\markcite{jackson}).  As in the general case, such a
transformation preserves the invariant interval $S_{\rm inv}$.  Unlike
rotations, it does not preserve the degree of polarization $|{\bf
S}|/S_{o}$ of the plane wave.  For example, there exists some
transformation that will completely depolarize partially polarized
radiation.  This is just the analogue of the statement in special
relativity that there always exists a reference frame in which two
events separated by a timelike interval occur at the same position in
space.  Equation \ref{eboost} may be represented in the basis of
equation \ref{pauli} as
\begin{equation}
{\bf b}_{\hat{\bf m}}(\beta) = \left(\begin{array}{cc}
              \cosh \beta + m_{q} \sinh \beta & (m_{u}-im_{v})\sinh \beta \\
              (m_{u}+im_{v})\sinh \beta & \cosh \beta - m_{q} \sinh \beta
              \end{array} \right)
\label{eboostbasis}
\end{equation}
where ${\hat {\bf m}} =(m_{\hat q}, m_{\hat u}, m_{\hat v})$.  The
boost of equation \ref{sboost} in this basis becomes\begin{equation}
{\bf B}_{\hat{\bf m}}(2\beta) = \left(\begin{array}{cccc} \cosh 2\beta
& m_{q}\sinh 2\beta & m_{u}\sinh 2\beta & m_{v}\sinh 2\beta \\
m_{q}\sinh 2\beta & 1+2m_{q}^{2}\sinh^{2} \beta & 2m_{q}m_{u}\sinh^{2}
\beta & 2m_{q}m_{v}\sinh^{2} \beta \\ m_{u}\sinh 2\beta &
2m_{q}m_{u}\sinh^{2} \beta & 1+2m_{u}^{2}\sinh^{2} \beta &
2m_{u}m_{v}\sinh^{2} \beta \\ m_{v}\sinh 2\beta & 2m_{q}m_{v}\sinh^{2}
\beta & 2m_{u}m_{v}\sinh^{2} \beta & 1+2m_{v}^{2}\sinh^{2} \beta
\end{array}\right)
\label{sboostbasis}
\end{equation}
The rotations of equations \ref{erotbasis} and \ref{srotbasis} and
boosts of equation \ref{eboostbasis} and \ref{sboostbasis} constitute
a subset of the Lorentz transformations that suffices for the radio
astronomical applications presented in the next section.

Finally, systems consisting of multiple physical processes are readily
modelled through successive application of the above transformations.
Of particular use in the analysis of such composite systems are the
commutation relations
\begin{equation}
\left[{\bf R}_{\hat {\bf n}}(\alpha){\bf R}_{\hat {\bf n}}(\beta)\right] = 
\left[{\bf R}_{\hat {\bf n}}(\alpha){\bf B}_{\hat {\bf n}}(\beta)\right] = 
\left[{\bf B}_{\hat {\bf n}}(\alpha){\bf B}_{\hat {\bf n}}(\beta)\right] = 0
\label{commrels}
\end{equation}
As we shall see in the next section, these relationships are useful in
determining whether the transformations that arise from different
physical processes or instrumental elements commute with one another.

\section{The Propagation and Reception of Radio Waves}
\label{radiomodel}

Let us examine some practical situations in which linear, invertible
transformations of polarized radiation arise.  One such example is a
rotation of the electric field vector about the direction of
propagation.  This transformation may represent a rotation of a
physical device with respect to the plane wave.  Another example
arises from Faraday rotation, which occurs when radio waves propagate
through a magnetized plasma.  Such a transformation may be written as
${\bf r}_{\hat v}(\phi)$.  Equivalently, the transformation of the
Stokes parameters is ${\bf R}_{\hat v}(2\phi)$.  A similar
transformation generates a phase delay between the two components of
${\bf E}$, and may be written as the rotation ${\bf r}_{\hat q}(\psi)$
.  In optics a physical device that induces such a phase delay is
called a compensator.  This transformation arises in electronics when
two signals traverse different cable lengths.  Both ${\bf r}_{\hat
v}(\phi)$ and ${\bf r}_{\hat q}(\psi)$ are unitary, and preserve the
degree of polarization $|{\bf S}|/S_{o}$ and the invariant
interval $S_{\rm inv}$.

Differential amplification or attenuation of the components of ${\bf
E}$ provide examples of a non-unitary transformation.  Consider the
transformation
\begin{equation}
{\bf g} = \tg  
\end{equation}
We may write this as ${\bf g} = \sqrt{g_{a}g_{b}}\; {\bf b}_{\hat
q}(\beta)$, where $\beta = \ln(g_{b}/g_{a})$.  The Stokes parameters
transform as $g_{a}g_{b}{\bf B}_{\hat q}(2\beta)$.  Note that this
transformation does not preserve $|{\bf S}|/S_{o}$, and preserves
$S_{\rm inv}$ only up to the factor $g_{a}g_{b}$.

Next consider two orthogonal elliptically polarized waves with axial
ratio $\tan \chi$ and orientation $\theta$ (Chandrasekhar 1960).
\begin{equation}
\begin{array}{cc}
e_{a} = \left(\begin{array}{c}
	\cos \theta \cos \chi - i \sin \theta \sin \chi \\
	-\sin \theta \cos \chi - i \cos \theta \sin \chi
	\end{array}\right) & 
e_{b} = \left(\begin{array}{c}
	\sin \theta \cos \chi - i \cos \theta \sin \chi \\
	\cos \theta \cos \chi + i \sin \theta \sin \chi 
	\end{array}\right)
\end{array}
\end{equation}
We may perform a change of basis by forming a matrix ${\bf s}$ with
rows $e_{a}^{\dagger}$ and $e_{b}^{\dagger}$.  This matrix may be
factored as ${\bf s}(\theta, \chi) = {\bf r}_{\hat u}(\chi){\bf
r}_{\hat v}(\theta)$.  A transformation from a linear to a circular
basis occurs for $\chi = \pi/4$, and in the case where $\theta=\pi/4$
results simply in a cyclic permutation of the indices $({\hat q},
{\hat u}, {\hat v}) \rightarrow ({\hat v}, {\hat q}, {\hat u})$ in
equation \ref{pauli}.  The transformation ${\bf s(\theta, \chi)}$ may
also be regarded as representing a receiver with two receptors
sensitive to orthogonal forms of elliptical radiation.  The process of
reception constitutes a projection of ${\bf E}$ onto the receptors of
the receiver, which is represented by matrix multiplication of ${\bf
E}$ by ${\bf s(\theta, \chi)}$.  Clearly the choice of a rotation
about the $\hat{u}$ axis followed by one about the $\hat{v}$ axis is
not unique.  One alternative specification of elliptically polarized
radiation is given in terms of an orientation and phase delay as ${\bf
r}_{\hat v}(\theta){\bf r}_{\hat q}(\phi)$.  In optics this
transformation is accomplished through a device known as a Babinet
compensator (Born \& Wolf 1980).

Now let us consider a receiver sensitive to two forms of elliptical
polarization that are not necessarily orthogonal.  This situation may
arise in practice from imperfections in the construction of a receiver
(Conway \& Kronberg 1969\markcite{conway}, Stinebring et
al. 1984\markcite{stinebring}).  The transformation may be written as
\begin{equation}
{\bf c} = \cmatrix
\label{defc}
\end{equation}
where the two probes of the receiver are sensitive to elliptical
radiation with axial ratios $\chi_{a}, \chi_{b}$ and orientations
$\theta_{a}, \theta_{b}$.  For the case $\theta_{a} = \theta_{b}$ and
$\chi_{a} = \chi_{b}$ equation \ref{defc} simplifies to the unitary
transformation ${\bf s}(\theta,\chi)$.  In the general case such a
transformation does not conserve energy.  With the definitions
\begin{equation}
\begin{array}{cc}
\begin{array}{l}
\sigma_{\theta} = \theta_{a} + \theta_{b} \\
\delta_{\theta} = \theta_{a} - \theta_{b} \\
\end{array}
\begin{array}{l}
\sigma_{\chi} = \chi_{a} + \chi_{b} \\
\delta_{\chi} = \chi_{a} - \chi_{b} \\
\end{array}
\end{array}
\end{equation}
we may write ${\bf c} = {\bf c}'{\bf r}_{\hat v}(\sigma_{\theta}/2)$, where
\begin{equation}
{\bf c}' = \left(\begin{array}{cc}
	\cos (\sigma_{\chi}/2+\delta_{\chi}/2) \cos \delta_{\theta}/2 + & 
	-\cos (\sigma_{\chi}/2+\delta_{\chi}/2) \sin \delta_{\theta}/2 + \\
	i\sin (\sigma_{\chi}/2+\delta_{\chi}/2) \sin \delta_{\theta}/2 & 
	i\sin (\sigma_{\chi}/2+\delta_{\chi}/2) \cos \delta_{\theta}/2 \\ 
	-\cos (\sigma_{\chi}/2-\delta_{\chi}/2) \sin \delta_{\theta}/2 + & 
	\cos (\sigma_{\chi}/2-\delta_{\chi}/2) \cos \delta_{\theta}/2 + \\
	i \sin (\sigma_{\chi}/2-\delta_{\chi}/2) \cos \delta_{\theta}/2 & 
	i\sin (\sigma_{\chi}/2-\delta_{\chi}/2) \sin \delta_{\theta}/2 
	\end{array}\right)
\end{equation}
For a nearly orthogonal receiver with receptors sensitive to linearly
polarized radiation, this matrix may be written to first order in
these parameters as ${\bf c}' = {\bf b}_{\hat
u}^{(1)}(\delta_{\theta}/2){\bf b}^{(1)}_{\hat v}(\delta_{\chi}/2){\bf
r}_{\hat u}^{(1)}(\sigma_{\chi}/2)$.  Here the superscript $(1)$
indicates that these transformations are first order in their
arguments.  

These examples may be combined to form a model for the propagation and
reception of radio waves.
\begin{equation}
{\bf t} = \sqrt{g_{a}g_{b}} {\bf b}_{\hat{q}}(\beta)
                {\bf r}_{\hat{q}}(\Phi_{I})
                {\bf b}_{\hat u}^{(1)}(\delta_{\theta}/2)
		{\bf b}_{\hat v}^{(1)}(\delta_{\chi}/2)
		{\bf r}_{\hat u}^{(1)}(\sigma_{\chi}/2)
		{\bf r}_{\hat v}(\sigma_{\theta}/2)
		{\bf r}_{\hat{v}}(\zeta)
                {\bf r}_{\hat{v}}(\Phi_{\rm iono})
                {\bf r}_{\hat{v}}(\Phi_{\rm ISM})
\label{emodel}
\end{equation}
where $\Phi_{\rm iono}$ and $\Phi_{\rm ISM}$ are the angles arising from
Faraday rotation in the ionosphere and the interstellar medium,
respectively, $\zeta$ is the angle between the frame of the receiver
and that of the sky, and ${\Phi_{I}}$ is the instrumental phase delay
arising from differing electronic pathlengths.  The equivalent
transformation law for the Stokes parameters is simply obtained from
equation \ref{emodel}.
\begin{equation}
S' = g_{a}g_{b} {\bf B}_{\hat{q}}(2\beta)
                {\bf R}_{\hat{q}}(2\Phi_{I})
                {\bf B}_{\hat u}^{(1)}(\delta_{\theta})
		{\bf B}_{\hat v}^{(1)}(\delta_{\chi})
		{\bf R}_{\hat u}^{(1)}(\sigma_{\chi})
		{\bf R}_{\hat v}(\sigma_{\theta})
                {\bf R}_{\hat{v}}(2\zeta)
                {\bf R}_{\hat{v}}(2\Phi_{\rm iono})
                {\bf R}_{\hat{v}}(2\Phi_{\rm ISM})\;S
\label{smodel}
\end{equation}
The analysis of this model is simplified through the commutation
relations in equation \ref{commrels}.  Amplifications and phase delays
in the downconversion chain represent boosts and rotations with
respect to the $\hat{q}$ axis, so it is easy to see that the order of
amplifiers and relative electronics delays does not matter.  Terms
from individual components may simply be collected into the overall
parameters $\beta$ and $\Phi_{I}$.  Similarly, rotations about the
same axis commute, and rotations about the $\hat{v}$ axis by the
angles $\zeta$, $\Phi_{\rm iono}$, and $\Phi_{\rm ISM}$ all have the
same signature.  Note that each of the parameters in this model may
have its own time and frequency dependence.  For example, $\Phi_{\rm
iono}$ fluctuates in time as the ionospheric column density changes,
and scales as $\nu^{-2}$ from the cold plasma dispersion relation.
Finally, calibration of a polarimetric observation is accomplished
through the inversion of equations \ref{emodel} or \ref{smodel}.
Naturally such an inversion requires a knowledge of the parameters in
this model.

\section{Discussion}
\label{discussion}

The representation of polarimetric transformations in terms of the
Lorentz group provide a simple context in which to analyze polarized
radiation.  This formalism is particularly relevant for the
calibration of polarimetric data, and greatly simplifies the
discussion from a qualitative standpoint.  Propagation or instrumental
effects that give rise to the rotations of equation \ref{srotbasis}
change the polarized component of the radiation ${\bf S}$, thus
obscuring the properties of the true, emitted light.  Those effects
that take the form of a boost transformation mix the total intensity
$S_{o}$ and the polarized component of the radiation ${\bf S}$.
Such transformations can have a particularly detrimental effect on
polarimetric observations.  In many astrophysical applications $S_{o}$
is much larger than $|{\bf S}|$, so that even boosts nearly
equivalent to the identity matrix may completely corrupt the polarized
flux.  Observations that aim to detect very small polarized fractions,
such as the polarized component of the Cosmic Microwave Background
radiation or the circularly polarized radio emission from Active
Galactic Nuclei, are particularly vulnerable.  For applications that
require extremely high precision, the mixing of ${\bf S}$ into
$S_{o}$ can corrupt an observation.  One such example has been seen in
high-precision pulsar timing, where differential amplification and
receiver cross-talk induce time dependent mixing of the pulse
profiles, thereby modifying the total intensity profile and
systematically shifting the times of arrival (Britton \etal, in preparation).
For pulsar observations the invariant $S_{\rm inv}$ proves
particularly useful, since an invariant pulse profile may be formed
that is independent of propagation and instrumental effects.  This
invariant profile may then be used for timing or to investigate pulse
variability that may be intrinsic to the pulsar.

\acknowledgements I thank Ren\'{e} Grognard, Richard Manchester,
Geoffrey Opat, and Matthew Bailes for useful conversations.  I thank
the referee for pointing out recent literature on the application of the 
Lorentz group to optics.


\begin{references}
\reference{barakat}{Barakat, R. 1963, J. Opt. Soc. Am. 53, 317}

\reference{born}{Born, M. \& Wolf, E. 1980, ``Principles of Optics''
(Cambridge: University Press)}

\reference{brace}{Bracewell, R. N. 1986, ``The Fourier Transform and
its Applications'' (New York: McGraw Hill)}

\reference{BB}{Brown, C. S. \& Bak, A. E. 1995, Optical Engineering
34, 1625}

\reference{Chandra}{Chandrasekhar, S. 1960, ``Radiative Transfer''
(New York: Dover)}

\reference{cloude}{Cloude, S. R. 1986, Optik, 75, 26}

\reference{Conway}{Conway, R. G., \& Kronberg, P. P. 1969, MNRAS 142,
11}

\reference{fano}{Fano, U. 1954, Phys. Rev. 93, 121}

\reference{gstein}{Goldstien, H. 1980, ``Classical Mechanics''
(Reading, Massachusetts: Addison-Wesley)}

\reference{hknc}{Han, D., Kim, Y. S. \& Noz, M. E. 1997a,
Phys. Rev. E, 56, 6065}

\reference{jackson}{Jackson, J.D. 1975, ``Classical Electrodynamics''
(New York: John Wiley \& Sons)}

\reference{jones}{Jones, R. C. 1941, J. Opt. Soc. Am. 31, 488}

\reference{mueller}{Mueller, H. 1948, J. Opt. Soc. Am. 38, 661}

\reference{opat}{Opatrny, T. \& Perina, J 1993, Physics Letters A,
181, 199}

\reference{parrent}{Parrent, G. B. \& Roman, P. 1960, Nuovo Cimento,
15, 370}

\reference{pfb}{Pellat-Finet, P. \& Bausset, M. 1992, Optik, 90, 101}

\reference{linnets}{Ruston, H. \& Bordogna, J. 1966, ``Electric
Networks: functions, filters, analysis'' (New York: McGraw-Hill)}

\reference{stinebring}{Stinebring, D. R., Cordes, J. M., Rankin, J. M., 
Weisberg, J. M., \& Boriakoff, V. 1984, ApJS, 55, 247}

\reference{whitney}{Whitney, C. 1971, J. Opt. Soc. Am., 61, 1207}

\reference{wienera}{Wiener, N. 1930, Acta Math. 55, 118}

\reference{wolf}{Wolf, E. 1959, Nuovo Cimento, 13, 1165}

\reference{WuKi}{Tung, W. K. 1985, ``Group Theory in Physics''
(Philadelphia: World Scientific)}
\end{references}
\end{document}